\documentclass[12pt]{article}
\usepackage{amsmath,color}
\usepackage[dvips]{graphicx}
\input amssym

\def\red#1{{\color{red} #1}}

\def\red#1{{\color{red} #1}}

\def\red{{}}
\begin{document}

%%%%%%%%%%%%%%%%%%%%%%%%%%%%%%%%%%%%%%%%%%%%%%%%%%%%%%%%%%%%%%%%%%%%%
\def\inn{\,\rfloor\,}                  \def\bt{{\bar\tau}}
\def\br{{\bar\rho}}
\def\prg#1{\medskip\noindent{\bf #1}}  \def\ra{\rightarrow}
\def\lra{\leftrightarrow}              \def\Ra{\Rightarrow}
\def\nin{\noindent}                    \def\pd{\partial}
\def\dis{\displaystyle}                \def\inn{\,\rfloor\,}
\def\grl{{GR$_\Lambda$}}               \def\Lra{{\Leftrightarrow}}
\def\cs{{\scriptstyle\rm CS}}          \def\ads3{{\rm AdS$_3$}}
\def\Leff{\hbox{$\mit\L_{\hspace{.6pt}\rm eff}\,$}}
\def\bull{\raise.25ex\hbox{\vrule height.8ex width.8ex}}
\def\ric{{(Ric)}}                      \def\tric{{(\widetilde{Ric})}}
\def\tmgl{\hbox{TMG$_\Lambda$}}
\def\Lie{{\cal L}\hspace{-.7em}\raise.25ex\hbox{--}\hspace{.2em}}
\def\sS{\hspace{2pt}S\hspace{-0.83em}\diagup}   \def\hd{{^\star}}
\def\dis{\displaystyle}     \def\ul#1{\underline{#1}}
\def\pgt{{\scriptstyle\rm PGT}}
\def\ul{{\underline}}

%% \hook is a better version of \rfloor
\def\hook{\hbox{\vrule height0pt width4pt depth0.3pt
\vrule height7pt width0.3pt depth0.3pt
\vrule height0pt width2pt depth0pt}\hspace{0.8pt}}
\def\semidirect{\;{\rlap{$\supset$}\times}\;}
\def\first{\rm (1ST)}       \def\second{\hspace{-1cm}\rm (2ND)}
\def\bm#1{\hbox{{\boldmath $#1$}}}      \def\un#1{\underline{#1}}
\def\nb#1{\marginpar{{\large\bf #1}}}
\def\diag{{\rm diag}\,}

\def\G{\Gamma}        \def\S{\Sigma}        \def\L{{\mit\Lambda}}
\def\D{\Delta}        \def\Th{\Theta}
\def\a{\alpha}        \def\b{\beta}         \def\g{\gamma}
\def\d{\delta}        \def\m{\mu}           \def\n{\nu}
\def\th{\theta}       \def\k{\kappa}        \def\l{\lambda}
\def\vphi{\varphi}    \def\ve{\varepsilon}  \def\p{\pi}
\def\r{\rho}          \def\Om{\Omega}       \def\om{\omega}
\def\s{\sigma}        \def\t{\tau}          \def\eps{\epsilon}
\def\nab{\nabla}      \def\btz{{\rm BTZ}}   \def\heps{\hat\eps}
\def\bu{{\bar u}}     \def\bv{{\bar v}}     \def\bs{{\bar s}}
\def\te{{\tilde e}}   \def\tk{{\tilde k}}
\def\bare{{\bar e}}   \def\bark{{\bar k}}
\def\barom{{\bar\omega}}  \def\barg{{\bar g}}
\def\tphi{{\tilde\vphi}}  \def\tt{{\tilde t}}
\def\te{{\tilde e}}

\def\tG{{\tilde G}}   \def\cF{{\cal F}}    \def\bH{{\bar H}}
\def\cL{{\cal L}}     \def\cM{{\cal M }}   \def\cE{{\cal E}}
\def\cH{{\cal H}}     \def\hcH{\hat{\cH}}
\def\cK{{\cal K}}     \def\hcK{\hat{\cK}}  \def\cT{{\cal T}}
\def\cO{{\cal O}}     \def\hcO{\hat{\cal O}} \def\cV{{\cal V}}
\def\tom{{\tilde\omega}}  \def\cE{{\cal E}}     \def\tR{{\tilde R}}
\def\cR{{\cal R}}    \def\hR{{\hat R}{}}   \def\hL{{\hat\L}}
\def\tb{{\tilde b}}  \def\tA{{\tilde A}}   \def\tv{{\tilde v}}
\def\tT{{\tilde T}}  \def\tR{{\tilde R}}   \def\tcL{{\tilde\cL}}
\def\hy{{\hat y}}    \def\tcO{{\tilde\cO}} \def\hom{{\hat\om}}
\def\he{{\hat e}}
%%%%%%%
\def\nn{\nonumber}                    \def\vsm{\vspace{-10pt}}
\def\be{\begin{equation}}             \def\ee{\end{equation}}
\def\ba#1{\begin{array}{#1}}          \def\ea{\end{array}}
\def\bea{\begin{eqnarray} }           \def\eea{\end{eqnarray} }
\def\beann{\begin{eqnarray*} }        \def\eeann{\end{eqnarray*} }
\def\beal{\begin{eqalign}}            \def\eeal{\end{eqalign}}
\def\lab#1{\label{eq:#1}}             \def\eq#1{(\ref{eq:#1})}
\def\bsubeq{\begin{subequations}}     \def\esubeq{\end{subequations}}
\def\bitem{\begin{itemize}}           \def\eitem{\end{itemize}}
\renewcommand{\theequation}{\thesection.\arabic{equation}}

%%%%%%%%%%%%%%%%%%%%%%%%%%%%%%%%%%%%%%%%%%%%%%%%%%%%%%%%%%%%%%%%%%%%%%%%%%
\title{5D Lovelock gravity: new exact solutions with torsion}
\author{ B. Cvetkovi\'c and D. Simi\'c\footnote{
        Email addresses: {cbranislav@ipb.ac.rs, dsimic@ipb.ac.rs}}\\
Institute of Physics, University of Belgrade \\
                      Pregrevica 118, 11080 Belgrade, Serbia}
\date{\today}
\maketitle

\begin{abstract}
Five-dimensional Lovelock gravity is investigated in the first order
formalism. A new class of exact solutions is constructed: the BTZ
black rings  with and without torsion. We show that our solution
with torsion exists in a different sector of the Lovelock gravity,
as compared to the Lovelock Chern-Simons sector or the one
investigated by Canfora et al. The conserved charges of the solutions
are found using Nester's formula, and the results are confirmed by
the canonical method. We show that the theory linearized around the
background with torsion possesses two additional degrees of freedom
with respect to general relativity. 	
\end{abstract}

\section{Introduction}

General theory of relativity introduced a revolution in our
understanding of space-time and gravity, the influence of which on
modern physics can hardly be emphasized enough---almost all
present investigations in high-energy physics are, in certain
way, related to it. On one hand, General theory of relativity
has been very successful in explaining experimental results,
but on the other, it produced a lot of problems for physicists to
solve. The first of them is the problem of singularities, appearing
quite often in gravitational solutions; there are theorems which show
that singularities must appear under certain physically reasonable
assumptions \cite{x4}. This situation inspired research in the
direction of alternative theories of gravity, with an idea of
finding a singularity free theory that reproduces experimental
results equally well as general relativity.

The second problem is quantization of general theory of relativity.
Inability to quantize general relativity in a standard way, like
Yang-Mills theories, motivated physicists to search for alternatives, on
one side for different quantization procedure (LQG) and on the other
for modifications of the original theory (extra dimensions,
SUSY, string theory, alternative theories of gravity)
\cite{2}\cite{7}\cite{8}\cite{9}. In this paper we shall focus on an
alternative theory of gravity with one extra dimension -- Lovelock gravity
in five dimensions (5D).

Lovelock gravity is one of many generalizations of general relativity,
physically appealing because of its similarity to the former. It
possesses equations of motion which are the second order
differential equations, it is ghost free etc. But beyond this, most
of its basic properties are not well known, and as the old saying says
"Devil is in the details". Firstly, not many solutions are known and those
constructed usually are torsionless or belong to some special point in
the parameter space \cite{12,13,18,19,y1}. Secondly, symmetries and
local degrees of freedom of the theory are not known for the generic
choice of parameters, but only for the special case of
Lovelock-Chern-Simons gravity \cite{5}.

In this paper we shall introduce new solutions with(out) torsion within
Lovelock gravity in 5D by using the first order formulation. The
most interesting of them  are the BTZ black rings with(out) torsion,
the properties of which can be analyzed by using the canonical
formalism. The canonical analysis is a powerful tool for studying
gauge theories, but it is not limited solely to them. It gives
a well-defined procedure for determining symmetries of a theory,
construction of the symmetry generators, and for counting the
number of local degrees of freedom. Applying the canonical analysis
to a theory is extremely rewarding because of the already
mentioned results it gives. Note, in particular, that the most
reliable approach to conserved charges in gravity is based on the
canonical analysis \cite{6,10}. The main aspect of this approach
consists in demanding the canonical generators to have well-defined
functional derivatives. For a given asymptotic behavior of the
fields, this condition usually requires the form of the generators to be
improved by adding suitable surface terms.

The paper is organized as follows. Section 2 contains a short review of
Poincar\'e gauge theory of gravity and Lovelock gravity. Third section is
devoted to the new solutions of 5D Lovelock gravity -- the BTZ black
rings with(out) torsion. The conserved charges for these solutions
are computed by using Nester formula \cite{x23}. In  section 4,  we
construct the canonical  generator of gauge transformations,
local translations and Lorentz rotations, and compute the
canonical conserved charges for the solutions constructed in section 3,
confirming the results obtained in section 3. In section 5, we
investigate canonical structure of the theory linearized around the
solution with torsion,  and conclude that in this sector, the theory
exhibits additional degrees of freedom.

Our conventions are given by the following rules: the Latin
indices refer to the local Lorentz frame, the Greek indices refer
to the coordinate frame; the first letters of both alphabets
$(a,b,c,...;$ $\a,\b,\g,...)$ run over $1,2,\dots D-1$ the middle alphabet
letters $(i,j,k,...;\m,\n,\l,...)$ run over $0,1,2,\dots D-1$; the signature
of spacetime is $\eta=(+,-,\dots,-)$; totally antisymmetric tensor
$\ve^{i_1i_2\dots i_D}$ and the related tensor density $\ve^{\m_1\mu_2\dots\m_D}$ are both
normalized so that $\ve^{01\dots D-1}=1$. \red{The symbol "$\wedge$"  of exterior (wedge) product between forms  is omitted
for simplicity}.

\section{Lovelock gravity}

\prg{PGT in brief.} The basic gravitational variables in PGT are the
vielbein $e^i$ and the Lorentz connection $\om^{ij}=-\om^{ji}$
(1-forms). The field strengths corresponding to the gauge potentials
$e^i$ and $\om^{ij}$ are the torsion $T^i$ and the curvature $R^{ij}$
(2-forms): $T^i= de^i+\om^i{_m}\wedge e^m$
$R^{ij}=d\om^{ij}+\om^i{_m}\wedge \om^{mj}$. Gauge symmetries of the theory
are local translations and local Lorentz rotations, parametrized by
$\xi^\m$ and $\ve^{ij}$.

In local coordinates $x^\m$, we can expand the vielbein and the
connection 1-forms as $e^i=e^i{_\m}dx^\m$, $\om^i=\om^i{}_\m dx^\m$.
Gauge transformation laws have the form
\bea
\d_0 {\red e}^i{_\m}&=& \ve^{ij}{\red e}_{j\m}-(\pd_\m\xi^\r){\red e}^i{_\r}
     -\xi^\r\pd_\r{\red e}^i{}_\m=:\d_\pgt {\red e}^i{}_\m\, ,      \nn\\
\d_0\om^{\red ij}{_\m}&=&\nab_\m\ve^{ij}
     -(\pd_\m\xi^\r)\om^{\red ij}{_\r}
     -\xi^\r\pd_\r\om^{\red ij}{}_\m=:\d_\pgt\om^{\red ij}{}_\m\, ,    \lab{2.1}
\eea
and the field strengths are given as
\bea
&&T^i=\nabla {\red e}^i\equiv d{\red e}^i+\om^{\red ij}\wedge {\red e}_j
     =\frac{1}{2}T^i{}_{\m\n}dx^\m\wedge dx^\n\, ,         \nn\\
&&R^{ij}=d\om^{ij}+\om^{ik}\wedge\om_k{^j}
      =\frac{1}{2}R^{ij}{}_{\m\n}dx^\m\wedge dx^\n\, ,        \lab{2.2}
\eea
where $\nabla=dx^\m\nabla_\m$ is the covariant derivative.

To clarify the geometric meaning of the above structure, we introduce
the metric tensor as a specific, bilinear combination of the \red{vielbeins}
\bea
&&g=\eta_{ij}e^i\otimes e^j=g_{\m\n}dx^\m\otimes dx^\n \, ,\nn\\
&& g_{\m\n}=\eta_{ij}e^i{_\m}e^j{_\n}\, ,
            \qquad \eta_{ij}=(+,-,-,\red{-,-})\, .                   \nn
\eea
Although metric and connection are in general independent
dynamical/geometric variables, the antisymmetry of $\om^{ij}$ in PGT is
equivalent to the so-called {\it metricity condition\/}, $\nabla
g=0$. The geometry whose connection is restricted by the metricity
condition (metric-compatible connection) is called {\it
Riemann-Cartan geometry\/}. Thus, PGT has the geometric structure of
Riemann-Cartan space.

The connection $\om^{ij}$ determines the parallel transport in the local
Lorentz basis. Being a true geometric operation, parallel transport is
independent of the basis. This property is incorporated into PGT via
the so-called {\it vielbein postulate\/}, \red{the vanishing of the total covariant derivative
of $e^i{_\mu}$
$$
D_\m(\om+\G)e^i{_\nu}:=\pd_\m e^i{_\n}+\om^{ij}{}_\m e_{j\n}-\G^\r{}_{\n\m}e^i{_\r}=0\,,
$$
where $\G^\r{}_{\n\m}$ is affine connection and torsion is defined by $T^\r{}_{\m\n}=\G^\r{}_{\n\m}-\G^\r{}_{\m\n}$.
Previous relation}  implies the identity
\be
\om_{ijk}=\D_{ijk}+K_{ijk}\, ,                               \lab{2.3}
\ee
where $\D$ is Riemannian (Levi-Civita) connection, and
$K_{ijk}=-\frac{1}{2}(T_{ijk}-T_{kij}+T_{jki})$ is the contortion. \red{Latin indices are changed into Greek and vice versa
by means of vielbeins (and its inverse). Namely, $X^i=e^i{_\m}X_\mu$ and $X^\mu=e_i{^\m}X^i$.}
\red{For the details see \cite{10}.}

\prg{Lovelock action and equations of motion.} Lovelock gravity can also be considered in the framework of PGT.
Dimensionally continued Euler density $L_p$ in $D$ dimensions is defined as
\be
L_p=\ve_{i_1 i_2 \dots i_D} R^{i_1i_2} \dots R^{i_{2p-1} i_{2p}}e^{i_{2p+1}} \dots e^{i_D},
\ee
where  $p$ is the number of curvature tensors in Euler density. In the
previous relation we omitted the wedge product for simplicity. General
form of the Lovelock gravity Lagrangian \cite{1} in 5$D$ is a linear
combination of all dimensionally continued Euler densities in 5
dimensions:
\bsubeq\lab{2.5}
\bea
I=\frac{\a_0}5I_0+\frac{\a_1}3I_1+\a_2 I_2\,,
\eea
where:
\bea
&&I_0=\int\ve_{ijkln}e^{i}e^{j}e^{k}e^{l}e^{n}\,,\nn\\
&&I_1=\int\ve_{ijkln}R^{ij}e^{k}e^{l}e^{n}\,,\nn\\
&&I_2=\int\ve_{ijkln}R^{ij}R^{kl}e^{n}\,.
\eea
\esubeq

\prg{Field equations.}
Variation of the action with respect to vielbein $e^i$
and connection $\om^{ij}$ yields the gravitational field equations:
\bsubeq\lab{2.7}
\bea
&&\ve_{ijkln}\left(\a_0e^{j}e^{k}e^{l}e^{n}+\a_1R^{jk}e^le^n+\a_2R^{jk}R^{ln}\right)=0\,,\lab{2.7a}\\
&&\ve_{ijkln}\left(\a_1 e^ke^l+2\a_2R^{kl}\right)T^n=0\,.\lab{2.7b}
\eea
\esubeq
Let us note that in the generic case, the field equations \eq{2.7}
imply that torsion can be non-vanishing.

For later convenience, let us present the tensor form of the
field equations:
\bsubeq\lab{2.8}
\bea
&&\ve^{\m\n\r\s\t}_{ijkln}\left(\a_0e^j{_\n}e^k{_\r}e^l{_\s}e^n{_\t}+\frac12\a_1R^{jk}{}_{\n\r}e^l{_\s}e^n{_\t}+\frac14\a_2R^{jk}{}_{\n\r}R^{ln}{}_{\s\t}\right)=0\,,\\
&&\ve^{\m\n\r\s\t}_{ijkln}\left(\a_1 e^k{_\n}e^l{_\r}+\a_2R^{kl}{}_{\n\r}\right)T^n{}_{\s\t}=0\,,
\eea
\esubeq
where $\ve^{\m\n\r\s\t}_{ijkln}:=\ve^{\m\n\r\s\t}\ve_{ijkln}$.

\prg{Consequences of field equations.} If we take covariant derivative of the \eq{2.7a} make use of the Bianchi identities and multiply \eq{2.7b} with $e^j$ we get the following system:
\bea
&&\ve_{ijkln}\left(2\a_0e^je^ke^l+\a_1R^{jk}e^l\right)T^n=0\,,\nn\\
&&\ve_{ijkln}\left(\a_1e^je^ke^l+2\a_2R^{jk}e^l\right)T^n=0\,.\nn
\eea
In the case $4\a_0\a_2-\a_1^2\neq 0$ the previous set of equations reduces to the the following conditions:
\bsubeq
\bea
&&v_i:=T^j{}_{ji}=0\,,\\
&&R^{jk}{}_{ir}T^r{}_{jk}-2Ric^j{_k}T^k{}_{ij}=0\,,
\eea
\red{where $Ric^j{_k}:=R^{jl}{}_{kl}$ is Ricci tensor.}
\esubeq

Therefore, in the generic case torsion is {\it traceless} and \red{second irreducible component of torsion ${}^{(2)}T_i$ vanishes. For details
on irreducible decomposition of torsion and curvature in PGT see \cite{yo}.}
Let us note that the condition $4\a_0\a_2-\a_1^2\neq 0$
is violated in the case of Lovelock Chern-Simons gravity.

\prg{Maximally symmetric solution.} The field equation admit the existence of maximally symmetric Riemannian solution (maximally symmetric Riemannian background) defined by:
\be
\bar R^{ij}=-\L e^ie^j\,,\quad \bar T^i=0\,,
\ee
where $\L$ is the effective cosmological constant iff:
\be
\a_0-\a_1\L+\a_2\L^2=0\,.
\ee
This equation can be solved for $\L$:
\be
\L_{\pm}=\frac{\a_1\pm\sqrt{\a_1^2-4\a_0\a_2}}{2\a_2}\,.
\ee
The solution is unique for $\a_1^2-4\a_0\a_2=0$, which is the case in Lovelock Chern-Simons gravity.

Let us note that in terms of $\L_{\pm}$ equations of motion \eq{2.7} take an elegant form:
\bsubeq\lab{2.12}
\bea
&&\ve_{ijkln}\left(R^{jk}+\L_+e^je^k\right)\left(R^{ln}+\L_-e^le^n\right)=0\,,\lab{2.12a}\\
&&\ve_{ijkln}\left(R^{kl}+\frac{\L_++\L_-}2e^ke^l\right)T^n=0\lab{2.12b}\,.
\eea
\esubeq
In obtaining these equations we assumed that $\a_2\neq 0$, and this
condition will be used in the rest of the paper,  because for
$\a_2=0$  the theory reduces to general relativity.

\section{New class of solutions}
\setcounter{equation}{0}

The search for a new class of solutions is inspired by Canfora et
al. \cite{3}, who found a solution of the type ${\rm AdS_2}\times{\rm S}_3$ when the coupling constants satisfy the relation
\be
\a_1^2=12\a_0\a_2\,,
\ee
which is different from the one satisfied in Lovelock Chern-Simons
gravity. We shall now construct another class of solutions of the
"complementary" type $\varSigma_3 \times \varGamma_2$, where $\varSigma_3$
and $\varGamma_2$ are three- and two-dimensional manifolds,
determined by solving the equations of motion. We start from the
following anzatz for curvature
\bea\lab{3.2}
&&R^{ab}=Ae^ae^b\,,\nn\\
&&R^{3a}=R^{4a}=0\,,\nn\\
&&R^{34}=Be^3e^4\,,
\eea
and torsion:
\bea\lab{3.3}
&&T^a=p\ve^{abc}e_be_c\,,\nn\\
&&T^3=T^4=0\,.
\eea
In the anzatz we used the following notation: $a,b,c,\dots\in\{0,1,2\}$
and $\ve^{abc}:=\ve^{abc34}$, and $A$, $B$ and $p$ are some
functions restricted by the equations of motion. Note that torsion is
totally antisymmetric and thus only \red{third irreducible component} ${}^{(3)}T^i$  is
non-vanishing, \red{see \cite{yo}}. Let us now check whether anzatz solves the equations of
motion \eq{2.12}. From \eq{2.12b} we obtain:
$$
\left(B+\frac{\L_-+\L_+}2\right)p=0\,.
$$
Thus, one can have a vanishing torsion for $p=0$, or a non-vanishing
torsion for
\be
B=-\frac{\L_-+\L_+}{2},.\lab{3.4}
\ee
From \eq{2.12a} we obtain:
\bea
&&A(\L_-+\L_+)+2\L_-\L_+=0\,,\lab{3.5}\\
&&4\L_-\L_++(A+B)(\L_-+\L_+)+2AB=0\,.\lab{3.6}
\eea
If $\L_-+\L_+=0$, which is equivalent to $\a_1=0$, Eq. \eq{3.5}
implies $\a_0=0$ whereas $A$ remains undetermined;
otherwise, for $\a_1\neq 0$, we have
\be
A=-\frac{2\L_-\L_+}{\L_-+\L+}
\ee

Let us first analyze the case with non-vanishing torsion and
$\a_1\neq 0$, when $A$ and $B$ are both determined. By combining
equations \eq{3.4}, \eq{3.5}, \eq{3.6} and using Vieta's formulas:
$\L_-+\L_+=\frac{\a_1}{\a_2}$ and $\L_-\L_+=\frac{\a_0}{\a_2}$ we obtain
that solution exists in the sector:
\be
\a_1^2=8\a_0\a_2\,.\lab{3.8}
\ee
This sector is different from the one in \cite{3} and the
above solution is the first one in this sector. Using  Eqs.
\eq{3.4}, \eq{3.7} and \eq{3.8} we obtain

\be\lab{3.9}
A=\frac{B}{2}.
\ee

Now we turn to the  solution with vanishing torsion and $\a_1\neq 0$. In
this case $A$ is determined and $B$ is arbitrary, which can be used
to insure the validity of \eq{3.6}, which takes the form

\be
2\frac{\a_0}{\a_2}+B\left(\frac{\a_1}{\a_2}-4\frac{\a_0}{\a_1}\right)=0.\lab{3.7}
\ee
We see that if $\a_1^2-4\a_0\a_2=0$, which is the Lovelock Chern-Simons
gravity, for the validity of \eq{3.7}, one must have $\a_0=0$.
These two conditions imply $\a_1=0$, which is in contradiction
with our assumption; hence, the solution does not exist in the
Lovelock-Chern-Simons case. If $\a_1^2-4\a_0\a_2\neq0$ and $\a_1\neq
0$ (recall that we are not interested in General relativity so $\a_2\neq0$
also) we can choose any value of parameters obeying this conditions and
get a solution. So this class of solutions exist generically i.e. for
almost any choice of parameters.

For clarity of the exposure, we devote next few sections to
the most interesting solutions which belong to the class derived in this
section.

\subsection{BTZ black  ring with torsion}

For this case, the curvature takes the following form:
\bea\lab{3.11}
&&R^{ab}=qe^ae^b\,,\nn\\
&&R^{3a}=R^{4a}=0\,,\nn\\
&&R^{34}=-\frac{1}{r_0^2}e^3e^4\,,
\eea
while the torsion is given by
\bea\lab{3.12}
&&T^a=p\ve^{abc}e_be_c\,,\nn\\
&&T^3=T^4=0\,.
\eea

The Bianchi identity implies that $p$ is \emph{constant}, and
the Riemannian curvature reads:
\bea\lab{3.13}
&&\tR^{ab}=\left(q+\frac{p^2}4\right)e^ae^b\,,\nn\\
&&\tR^{3a}=\tR^{4a}=0\,,\nn\\
&&\tR^{34}=-\frac{1}{r_0^2}e^3e^4\,,
\eea
Therefore, we can introduce the AdS$_3$ radius $\ell$ as
$$
\frac{1}{\ell^2}:=q+\frac{p^2}4\,.
$$
Identity \eq{3.9} implies the following relation
\be
\frac{1}{\ell^2}=-\frac{1}{2r_0^2}+\frac{p^2}4\,.
\ee

In the AdS$_3$ sector, the anzatz for curvature and torsion is
solved by the AdS$_3$ solution with torsion, as well as
by the Ba\~nados, Teitelboim, Zanelli (BTZ)  black hole \cite{14}
with torsion. In the latter, physically more appealing case, the 5D
vielbein reads:
\bsubeq\lab{3.15}
\bea
&&e^0=Ndt\,,\quad e^1=N^{-1}dr\,,\quad e^2=r(d\vphi+N_\vphi dt)\,,\nn\\
&&e^3=r_0d\theta\,,\quad e^4=r_0\sin\th d\chi\,,
\eea
where
$$N^2=-2m+\frac{r^2}{\ell^2}+\frac{j^2}{r^2}\,,\qquad N_\vphi=\frac{j}{r^2}\,,$$
where $m$ and $j$ are (dimensionless) parameters. The Cartan
connection is given by:
\bea
&&\om^{ab}=\tom^{ab}-\ve^{abc}\frac{p}2e_c\,,\nn\\
&&\tom^{01}=-\frac{r}{\ell^2}dt-\frac{j}rd\vphi\,,\quad \tom^{12}=Nd\vphi\,,\nn\\
&&\tom^{20}=N^{-1}\frac{j}{r^2}dr\,,\nn\\
&&\om^{34}=\tom^{34}=-\cos\th d\chi\,,
\eea
where \red{$\tom^{ij}$ is  Riemannian connection.}
\esubeq
Let us note that the coordinate ranges are:
$$-\infty<t<+\infty\,,\quad0\leq r<+\infty\,,\quad 0\leq\vphi\leq 2\pi\,,\quad 0\leq\th\leq\pi\,,
\quad 0\leq\chi\leq 2\pi\,.$$
\prg{Killing vectors.} The maximal number of Killing vectors
of the solution with field strengths \eq{3.11}, \eq{3.12} and
\eq{3.13}, is $9=6+3$, since the AdS$_3$ solution with(out) torsion
has 6 Killing vectors, see \cite{15}. The solution \eq{3.15} has {\it
five} Killing vectors, since the BTZ solution possesses 2 Killing
vectors. They are given by:
\bea\lab{3.16}
&&\xi^{(1)}=\ell\frac{\pd}{\pd t}\,,\qquad \xi^{(2)}=\frac{\pd}{\pd\vphi}\,,\qquad \xi^{(3)}=\frac{\pd}{\pd\chi}\,,\nn\\
&&\xi^{(4)}=\sin\chi\frac{\pd}{\pd\th}+\cot\th\cos\chi\frac{\pd}{\pd\chi}\,,\qquad \xi^{(5)}=\cos\chi\frac{\pd}{\pd\th}-\cot\th\sin\chi\frac{\pd}{\pd\chi}\,.
\eea

\subsection{Riemannian BTZ ring}

For this case, the curvature (Riemannian) takes the following form:
\bea
&&R^{ab}=\frac1{\ell^2}e^ae^b\,,\nn\\
&&R^{3a}=R^{4a}=0\,,\nn\\
&&R^{34}=-\frac{1}{r_0^2}e^3e^4\,,
\eea
while the torsion equals zero, $T^i=0$.

Let us note that since torsion is zero, there are no further constraints
on $B$, so that we can chose $B=-\frac1{r_0^2}$. In terms of
the action constants we get:
\be
\frac 1{\ell^2}=-\frac{2\a_0}{\a_1}\,,\qquad \frac{1}{r_0^2}=\frac{2\a_0\a_1}{\a_1^2-4\a_0\a_2}
\ee
The solution exists provided that $\a_0\a_1<0$ and $\a_1^2-4\a_0\a_2<0$.
Let us note  this solution does not solve equations of motion in Lovelock Chern-Simons gravity.

The vielbein fields and connection take the same form as in \eq{3.15} with $p=0$, while Killing vectors are identical and
given by \eq{3.16}.

\subsection{Conserved charges}
In order to compute conserved charges, we shall make use of Nester
formula. Let us denote the difference between any variable $X$ and its
reference value $\bar X$ by $\D X=X-\bar X$. In 5D, the boundary term $B$
is a 3-form. With a suitable set of boundary conditions for the fields,
the proper boundary term reads \cite{x23}:
\be
B=(\xi\inn b^i)\D \t_i + \D b^i(\xi\inn\bt_i)
  +\frac 12(\xi\inn\om^i{_j})\D\r_i{^j}
  +\frac 12\D\om^i{_j}(\xi\inn\br_i{^j})\, ,
\ee
\red{where $\xi$ is an asymptotically Killing vector, while $\t_i$ and $\r_{ij}$ are covariant momenta corresponding to torsion and curvature, respectively.
The covariant momenta for the Lovelock action \eq{2.5} are given by:
\bea
&&\t_i:=\frac{\pd L}{\pd T^i}=0\,,\\
&&\r_{ij}=\frac{\pd L}{\pd R^{ij}}=2\ve_{ijkln}\left(\frac {\a_1}{3}e^ke^l+2\a_2 R^{kl}\right)e^n\,.
\eea
Consequently, we obtain:
\bea
&&\r_{ab}=4\ve_{abc}\left(\a_1-\frac{2\a_2}{r_0^2}\right)e^ce^3e^4\,,\nn\\
&&\r_{a3}=2\ve_{abc}\left(\a_1+2\a_2q\right)e^be^ce^4=\a_1\ve_{abc}e^be^ce^4\,,\nn\\
&&\r_{a4}=2\ve_{abc}\left(\a_1+2\a_2q\right)e^be^ce^3=\a_1\ve_{abc}e^be^ce^3\,,\nn\\
&&\r_{34}=2\ve_{abc}\left(\frac{\a_1}3+2\a_2q\right)e^ae^be^c=-\frac{\a_1}3e^ae^be^c\,.
\eea}

In our calculations of the boundary integrals, we use the coordinates
$x^\m=(t,r,\vphi,\th,\chi)$. \red {The background configuration is the one defined by zero values of the solution parameters $m=0$ and $j=0$.}  For the solutions with Killing vectors
$\pd_t$ and $\pd_\vphi$, the conserved charges are energy and angular
momentum, respectively:
\bsubeq\lab{3.3}
\bea
&&E=\int_{\pd\S}B(\pd_t)
   =\int_{\pd\S} e^i{_t}\D \t_i + \D e^i \bt_{it}
        +\frac12\om^{ij}{}_t\D\r_{ij} + \frac12\D\om^{ij}\br_{ijt}\, ,             \lab{3.3a}\\
&&J=\int_{\pd\S}B(\pd_\vphi)
   =\int_{\pd\S} e^i{_\vphi}\D \t_i + \D e^i \bt_{i\vphi}
        +\frac12\om^{ij}{}_\vphi\D\r_i + \frac 12\D\om^{ij}\br_{ij\vphi}\, ,             \lab{3.3b}
\eea
\esubeq
where $\pd\S$ is a boundary $S^1\times S^2$, located at infinity, described
by coordinates $\vphi,\th,\chi$.

Thus, conserved charges for for the black ring with torsion
and the Riemannian black ring are given by:
\be
E=8\pi^2r_0^2\left(\a_1-\frac{2\a_2}{r_0^2}\right)m\,,\qquad J=8\pi^2r_0^2\left(\a_1-\frac{2\a_2}{r_0^2}\right)j\,. \lab{3.24}
\ee
Let us note that the solution with torsion exists in the sector
$\a^2_1=8\a_0\a_2$, where both conserved charges vanish.

\section{Canonical gauge generator}
\setcounter{equation}{0}

As an important step in our examination of the asymptotic structure of
spacetime, we are going to construct the canonical gauge generator,
which is our basic tool for studying asymptotic symmetries and
conserved charges of 5D Lovelock gravity.

\subsection{Hamiltonian and constraints}
\setcounter{equation}{0}

The best way to understand the dynamical content of gauge symmetries is to explore
the  canonical generator, which acts on the basic dynamical
variables via the Poisson bracket (PB) operation. To begin the canonical analysis, we rewrite the
action \eq{2.5} as
\bea
&&I={\red \int d^5x\cL}\nn\\
&&\cL=\ve^{\m\n\r\s\t}_{ijkln}\int_{\cM} d^5 x\left(\frac{\alpha_0}5e^i{_\m}e^j{_\n}e^k{_\r}e^l{_\s}+
\frac{\a_1}6R^{ij}{}_{\mu \nu}e^k{_\r}e^l{_\s}+\frac{\a_2}4R^{ij}{}_{\mu \nu}R^{kl}{}_{\r\s}\right)e^n{_\t}\,.
\eea
\prg{1.}
The basic Lagrangian variables
$(e^i{}_\m,\om^{ij}{}_\m)$ and the corresponding canonical momenta
$(\pi_i{}^\m,\pi_{ij}{}^\m)$ are related to each other through the set
of primary constraints:
\bea
&&\phi_i{}^0:=\pi_i{}^0\approx 0\, ,
       \qquad \phi_{ij}{}^0:=\pi_{ij}{}^0\approx 0\, ,\nn\\
&&\phi_i{}^\a:=\pi_i{}^\a\approx 0\, ,
\qquad\phi_{ij}{}^\a:=\pi_{ij}{}^\a-2\ve^{0\a\b\g\d}_{ijkln}\left(\frac{\a_1}3 e^k{_\b}e^l{_\g}+\alpha_{2}R^{kl}{}_{\b\g}\right)e^n{_\d}\approx 0\, .
\eea
The algebra of primary  constraints is displayed in appendix A.

\red{Since Lagrangian is linear in velocities the canonical Hamiltonian is defined by formula $\cH_c=-\cL(\dot e^i{_\m}=0,\dot\om^{ij}{}_\m=0)$.
It  is linear in unphysical variables}:
\bea
&&\cH_c=e^i{_0}\cH_i+\frac12\om^{ij}{}_0\cH_{ij}+\pd_\a D^\a\,,\nn\\
&&\cH_i=-\ve^{0\a\b\g\d}_{ijkln}\left(\a_0e^j{_\a}e^k{_\b}e^l{_\g}e^n{_\d}+
\frac12\a_1R^{jk}{}_{\a\b}e^l{_\g}e^n{_\d}+\frac14\a_2R^{jk}{}_{\a\b}R^{ln}{}_{\g\d}\right)\,,\nn\\
&&\cH_{ij}=-\ve^{0\a\b\g\d}_{ijkln}\left(\a_1 e^k{_\a}e^l{_\b}+\a_2R^{kl}{}_{\a\b}\right)T^n{}_{\g\d}\,,\nn\\
&&D^\a=\ve^{0\a\b\g\d}_{ijkln}\om^{ij}{}_0\left(\a_1 e^k{_\b}e^l{_\g}+\a_2R^{kl}{}_{\b\g}\right)e^n{_\d}\,.
\eea

\prg{2.} Going over to the total Hamiltonian,

\be
\cH_T=\cH_c+u^i{_\m}\phi_i{^\m}+\frac 12u^{ij}{}_\m\phi_{ij}{}^\m\,,
\ee
we find that the consistency conditions of the primary constraints $\pi_i{^0}$ and $\pi_{ij}^0$ yield
the secondary constraints:
\be
\cH_i\approx 0\,,\quad \cH_{ij}\approx 0\,.
\ee
Let us note that these constraints reduce to the $\mu=0$ components of the Lagrangian field
equations \eq{2.8}.

The consistency of the remaining primary constraints $\phi_i{}^\a$,
and $\phi_{ij}{}^\a$  leads to the relations for
multipliers $u^i{}_\b$ and $u^{ij}{}_\b$:
\bea\lab{4.6}
&&\ve^{0\a\b\g\d}_{ijkln}\left[\underline{R}^{jk}{}_{0\b}\left(\a_1e^l{_\g}e^n{_\d}+\a_2 R^{ln}{}_{\g\d}\right)+
\left(\a_1R^{jk}{}_{\b\g}+4\a_0e^j{_\b}e^k{_\g}\right)e^l{_0}e^n{_\d}\right]=0\,,\nn\\
&&\ve^{0\a\b\g\d}_{ijkln}\left[\underline{T}^k{}_{0\b}\left(\a_1e^l{_\g}e^n{_\d}+\a_2R^{ln}{}_{\g\d}\right)+
\a_2\underline{R}^{kl}{}_{0\b}T^n{}_{\g\d}+\a_1e^k{_0}e^l{_\b}T^n{}_{\g\d}\right]=0\,,
\eea
where $\underline{T}^i{}_{0\a}=T^i{}_{0\a}(\dot e^i{_\a}\ra u^i{_\a})$ and
$\underline{R}^{ij}{}_{0\a}=R^{ij}{}_{0\a}(\dot\om^{ij}{}_{\a}\ra u^{ij}{}_\a)$.
Using the Hamiltonian equations of motion $\dot e^i{}_\a=u^i{}_\b$ and
$\dot\om^{ij}{}_\a=u^{ij}{}_\a$  these
relations reduce to the $\mu=\a$ components of the Lagrangian field
equations \eq{2.8}.

\prg{Further consistency procedure.} Some of the relations \eq{4.6} can be solved in terms of the
multipliers $u^i{_\a}$ and $u^{ij}{_\a}$, while the others may lead to
ternary constraints, whose consistency has to be examined as well.
However, this procedure is extremely sensitive to the particular sector of
the theory as we shall illustrate in the next section  (for the pure Lovelock theory see \cite{11}).  The final form of the total
Hamiltonian is given by:
\bea
&&\cH_T=\bar{\cH}_T+u^i{_0}\pi_i{^0}+\frac12u^{ij}{}_0\pi_{ij}{}^0+\left(u\cdot\phi\right)\,,\nn\\
&&\bar{\cH_T}=e^i{_0}\bar\cH_i+\frac12\om^{ij}{}_0\bar\cH_{ij}+\pd_\a\bar D^\a\,,\nn\\
&&\bar\cH_i=\cH_i+(\bar u\cdot\phi)\,,\nn\\
&&\bar\cH_{ij}=\cH_{ij}+(\bar u\cdot\phi)\,,\nn\\
&&\bar D^\a=D^\a+(\bar u\cdot\phi)\,,
\eea
where by $(u\cdot\phi)$ we denoted terms stemming form the \emph{undetermined} multipliers and belonging to the set $(u^i{_\b},u^{ij}{}_\b)$
 and by $(\bar u\cdot\phi)$ we denoted terms stemming form the \emph{determined} multipliers belonging to the same set.
\subsection{Canonical generator and charges}
The sure symmetries of the theory are local translations and local Lorentz
rotations. The general form of the canonical generator of the local
Poincar\'e transformations constructed by the Castellani procedure
\cite{x25} is given by:
\bea
&&G=-G_1-G_2\,,\nn\\
&&G_1=\dot\xi^\r\left(e^i{_\r}\pi_i{^0}+\frac12\om^{ij}{}_\r\pi_{ij}^0\right)+
\xi^\r\left(e^i{_\r}\bar\cH_i+\frac12\om^{ij}{_\r}\bar\cH_{ij}+C_{PFC}\right)\,,\nn\\
&&G_2=\frac12\dot\ve^{ij}\pi_{ij}{}^0+\frac12\ve^{ij}(\bar\cH_{ij}+C_{PFC})\,,\nn\
\eea
where $C_{PFC}$ are terms proportional to sure primary first class constraints $(\pi_i{^0},\pi_{ij}{}^0)$.

The canonical generator acts on dynamical variables via the PB operation,
hence, it should have well-defined functional derivatives. In order to
ensure this property, we have to improve the form of $G$ by adding a
suitable surface term $\G$, such that $\tG= G + \G$ is a well-defined
canonical generator. In this process, the asymptotic conditions play a
crucial role, see for instance \cite{x24,16}. Though we didn’t construct
the exact form of the canonical generator it still allows us to compute
canonical charges for the solutions found in section 3. Namely, if we
adopt the general principle that the quantities that
vanish on shell have an arbitrary fast asymptotic decrease, we obtain
that the on shell variation of the generator takes the following form:
\bea
\d G(\xi^t=\ell,\xi^\vphi=1)\approx\d \G=-\ell\d E_c-\d J_c\,,
\eea
where
\be
E_c=8\pi^2 r_0^2\left(\a_1-\frac{2\a_2}{r_0^2}\right)m\,,\qquad J_c=8\pi^2 r_0^2\left(\a_1-\frac{2\a_2}{r_0^2}\right)j\,,
\ee
are the \emph{canonical} conserved charges, which are identical to
the expressions \eq{3.24}, obtained from the Nester
formula.
\section{Linearized theory}
\setcounter{equation}{0}

The canonical structure of the full non-linear theory crucially depends on
the relations \eq{4.6}, as we already mentioned in the previous
section. In order to get a deeper insight into the structure of the
Lovelock gravity in the sector $\a_1^2=8\a_0\a_2$ we shall consider the
theory linearized around around the BTZ black ring with torsion
\eq{3.15}. The linearization is based on the expansion of the basic
dynamical variables  $(e^i{_\m},\om^{ij}{}_\m)$ and the related
conjugate momenta $(\pi_i{^\m},\pi_{ij}{}^\m)$ denoted shortly by $Q_A$:
\bea
Q_A=\bar Q_A+\tilde Q_A\,,
\eea
where  $\bar Q_A$  refers to the background (solution \eq{3.15} with $m=j=0$ and $p\neq 0$), while $\tilde Q_A$ denotes small
excitations.

 From the linearized form of the 60 relations \eq{4.6} we conclude that
  out of $60=5\times 4+10\times 4$ multipliers
  $(\tilde{u}^i{_\a},\tilde{u}^{ij}{}_\a)$,  46 are determined, while
  among 14 remaining relations there are 12 new constraints (since two
  pairs of them are identical), whose
 explicit form is given by:
 \bsubeq
\bea\lab{5.2}
&&\a_1\tR^{24}{}_{r\chi}+\a_1\sin\th\tR^{23}{}_{r\th}+4\a_0r_0\sin\th\te^2{_r}\approx 0\,,\\
&&\a_1\tR^{14}{}_{\vphi\chi}+\a_1\sin\th\tR^{13}{}_{\vphi\th}+4\a_0r_0\sin\th\te^2{_r}\approx 0\,,\\
&&\frac{r^2}{\ell}\left(\a_1\tR^{14}{}_{r\chi}+\a_1\sin\th\tR^{13}{}_{r\th}+2\a_0r_0\sin\th \te^1{_r}\right)\nn\\
&&-\a_1\tR^{24}{}_{\vphi\chi}-\a_1\sin\th\tR^{23}{}_{\vphi\th}-2\a_0r_0\sin\th\te^2{_\vphi}\approx 0\,,
\eea
\esubeq
and
\bsubeq
\bea\lab{5.3}
&&\tT^4{}_{r\chi}+\sin\th\tT^3{}_{r\th}\approx 0\,,\\
&&p\left(\a_1r_0\left(\te^4{_\chi}+\sin\theta\te^3{_\th}\right)+2\a_2\tR^{34}{}_{\th\chi}\right)\approx 0\,,\\
&&\tT^4{}_{\vphi\chi}+\sin\th\tT^3{}_{\vphi\th}\approx 0\,,\\
&&\a_1\frac r\ell r_0\sin\th\tT^2{}_{r\th}-2p\left(\a_1 r_0\sin\th\te^0{_\th}+2\a_2\tR^{04}{}_{\th\chi}\right)\approx 0\,,\\
&&\a_1\frac r\ell r_0\tT^2{}_{r\chi}-2p\left(\a_1 r_0\sin\th\te^0{_\chi}-2\a_2\tR^{03}{}_{\th\chi}\right)\approx 0\,,\\
&&\a_1 r_0\tT^1{}_{\vphi\chi}+2pr\left(\a_1 r_0\te^0{_\chi}-2\a_2\tR^{03}{}_{\th\chi}\right)\approx 0\,,\\
&&\tR^{03}{}_{r\chi}\approx 0\,,\\
&&\tR^{02}{}_{\th\chi}\approx 0\,,\\
&&\tR^{01}{}_{\th\chi}\approx 0\,.
\eea
\esubeq
Let us denote 12 constraints \eq{5.2} and \eq{5.3} by $\tilde\psi_A$. The
consistency conditions of $\tilde\psi_A$ leads to determination of 12
additional multipliers thus finishing the consistency procedure. Thus,
out of 60 multipliers $(\tilde{u}^i{_\a},\tilde{u}^{ij}{}_\a)$, 58
are determined, while 2 remain undetermined. By using the PB algebra
from appendix A we find:
\bea
&&\{\tilde\phi_{12}{}^r,\tilde\phi_i{^\a}\}\approx 0\,,\qquad \{\tilde\phi_{12}{}^r,\tilde\phi_{ij}{}^\a\}\approx 0\,,\nn\\
&&\{\tilde\phi_{12}{}^r,\tilde\psi_A\}\approx 0\,,\nn\\
&&\{\tilde\phi_{12}{}^\vphi,\tilde\phi_i{^\a}\}\approx 0\,,\qquad \{\tilde\phi_{12}{}^\vphi,\tilde\phi_{ij}{}^\a\}\approx 0\,,\nn\\
&&\{\tilde\phi_{12}{}^\vphi,\tilde\psi_A\}\approx 0\,.\nn
\eea
The undetermined multipliers  correspond to the constraints
$\tilde\phi_{12}{}^r$ and $\tilde\phi_{12}{}^\vphi$  which are FC.
The final classification of constraints is given by:

\begin{center}
\begin{tabular}{|l|l|l|}
\hline\hline
&First class&Second class\\
Primary&$\tilde\phi_i{^0},\tilde\phi_{ij}{^0},\tilde\phi_{12}{}^r,\tilde\phi_{12}{}^\vphi$&$\tilde\phi_i{^\a},\tilde\phi_{ij}{^\a}\quad ij\neq 12 \wedge \a \neq r,\vphi$\\
Secondary&$\tilde{\bar\cH}_i,\tilde{\bar\cH}_{ij}$&$\tilde\psi_A$\\
\hline\hline
\end{tabular}\\
\vspace{0.3cm}
Table 1. Classification of constraints
\end{center}
In total there are $N_1=32$ FC constraints and $N_2=70$ SC constraints. The number of propagating degrees of freedom
in phase space is:
$$
N^*=2N-2N_1-N_2=150-64-70=14\,.
$$
In the configuration space there are 7 degrees of freedom, 5 of them
correspond to GR  in $D=5$ and two are additional degrees of freedom. The
presence of two primary  FC constraints
$\tilde\phi_{12}{}^r,\tilde\phi_{12}{}^\vphi$ implies that there is an
additional gauge symmetry in the theory, as a consequence of the
fact that variables $\tom^{12}{}_r$ and $\tom^{12}{}_\vphi$ do not appear
in the linearized equations of motion.

\section{Conclusion}

In this paper we found  a new class of solutions of Lovelock gravity
in 5D, in the first order formalism. The most interesting
solutions are the BTZ black rings with(out) torsion. It is
shown that the solution with torsion exists provided that the parameters
of the theory satisfy the relation $\a_1^2=8\a_0\a_2$. This sector of
the parameter space is different from the one of Lovelock
Chern-Simons gravity, as well as from the sector investigated by
Canfora et al \cite{3}. Restricting our attention to the basic
properties of the solutions, we calculated the values of
conserved charges by using Nester's formula and the canonical
method. The canonical structure of the theory linearized around the
background with torsion shows that there are two additional degrees
of freedom, compared to general relativity.

\section*{Acknowledgements}
We thank Milutin Blagojevi\'c for useful remarks and suggestions. This work was supported by the Serbian Science foundation, Serbia, grant 171031.
\appendix
\section{The algebra of constraints}

The structure of the PB algebra of constraints is an important
ingredient in the  analysis of the Hamiltonian consistency
conditions. Starting from the fundamental PB $\{e^i{_\mu},\pi_j{^\nu}\}=\delta^i_j \d^\n_\m\d(\mathbf{x-x'})$
and $\{\om^{ij}{}_\m,\pi_{kl}{}^\nu\}=2\d^{[i}_k\d^{j]}_l\d^\n_\m\d(\mathbf{x-x'})$  we find PB between primary constraints:
\bea
&&\{\phi_i{^\a},\phi_{jk}{}^\b\} = -2\ve_{ijkln}^{0\a\b\g\d}(\a_1e^l{_\g}e^n{_\d}+\a_2R^{ln}{}_{\g\d})\d,\nn\\
&&\{\phi_{ij}{}^\a ,\phi_{kl}^\b\} =-8\a_2\ve_{ijkln}^{0\a\b\g\d} T^n{}_{\g\d}\d\,.
\eea

\end{document}